\documentclass[12pt,fleqn]{article}
\usepackage[utf8]{inputenc}
\usepackage[german,english]{babel}

\begin{document}
\title{{The Projective Andoyer transformation \\ and 
 the connection between\\ the 4-D isotropic oscillator and Kepler systems}}
\author{Sebasti\'an Ferrer \\
{\small Dpto de Matem\'atica Aplicada, Universidad de Murcia, 30071 Espinardo, Spain}\medskip\\
November 12, 2010 }
\date{\empty}
\maketitle 
\begin{abstract} 
Extending to 4 degrees of freedom a symplectomorphism used in attitude dynamics it is shown in a direct way the connection between the 4-D isotropic harmonic oscillator and the 3-D Kepler systems. This approach made transparent that only when we refer to rectilinear solutions, the {\sl bilinear relation} defining the KS transformation is needed.
\end{abstract}
\maketitle
\section{Introduction}
More than 40 years after the KS transformation \cite{Kus64},  the
epitome of which might be  the work of Cordani \cite{Corda}, 
any proposal of a  new insight on the connection between the 4-D isotropic oscillator and
Kepler systems might  be taken as the opposite. Nevertheless, the way in which many authors still  
deal with the relation between these integrable systems, has pushed finally to take the risk and
write this letter. Our claim is that, in order to present the connection of these systems we do not need to go by concepts like embeddings, weak extended canonical transformation formalism \cite{Sti}, or the more recently used geometric algebraic approaches \cite{Kummer, viva94, Bart}. We came to understand what is explained here studying  the 4-D isotropic oscillator reductions \cite{Egea07a}. More precisely, it is shown in a direct 
 way that the  Kepler-Coulomb flow appears as a part of the 4-D isotropic oscillator flow. \par
The core of our approach goes back to a basic concept in 3-D dynamics {\sl `the instantaneous plane of motion'} and the associated nodal-polar variables, already used in planetary theories by Hill \cite{Hill1913} and much later in satellite theory (see Deprit \cite{Dep70}). Today some authors refer to them  as {\sl Whittaker transformation} \cite{Dep81},\cite{Yanguas}. Nevertheless, for reasons which are unknown to this author, this chart still continues to be ignored, versus the spherical variables, not only in vast field of mechanics but even in the special area of Hamiltonian astrodynamics. \par

Inspired in the 3-D Whittaker transformation, what we propose here is to consider part of the Andoyer angles (see (\ref{ESA6})) as the ones defining a `plane of motion' joint with a function of the distance, which lead to the Kepler system as part of the 4-D oscillator. This author is convinced that the
use of {\sl `polar-nodal variables'} in 4-D context  will be of great benefit also in quantum mechanics, molecular
dynamics, etc. 
\par
The 4-D isotropic oscillator is an integrable dynamical system defined by the parametric Hamiltonian function
\begin{equation}\label{O4G}
{\cal H}_\omega= \frac{1}{2}\sum_i^4(Q_i^2 +\, 
\omega\,q_i^2),
\end{equation}
where $\omega$ is a  parameter. There is a large and uninterrupted literature about this system 
going from Jauch and Hill \cite{Jauch} to Waldvogel \cite{Wal08}, including Moser  
 \cite{Moser} and Iwai \cite{Iwai},
due to the fact that it is one of the rare few examples of maximally superintegrable systems (see Fass\`o \cite{Fasso}). 
\section{ Starting with  Projective Euler  variables} We consider the  transformation:
${\cal PE}_F: (\rho,\phi,\theta,\psi)\rightarrow (q_1,q_2,q_3,q_4)$, dubbed as {\sl Projective Euler} variables, given by
\begin{eqnarray}\label{parametros}
\vspace{-0.3cm}
&&\hspace{-0.8cm}q_1=F({\rho})\, \sin\frac{\theta}{2} \cos\frac{\phi-\psi}{2},\,
q_3=F({\rho})\, \cos\frac{\theta}{2} \sin\frac{\phi+\psi}{2}, \\
&&\hspace{-0.85cm}q_2=F({\rho})\, \sin\frac{\theta}{2} \sin\frac{\phi-\psi}{2},\,
q_4=F({\rho})\,\cos\frac{\theta}{2} \cos\frac{\phi+\psi}{2},\nonumber
\end{eqnarray}
with $(\rho,\phi,\theta,\psi)\in R^+\times[0,2\pi)\times (0,\pi)\times
\left(-\frac{\pi}{2},\frac{\pi}{2}\right)$. When $F({\rho})=1$, the transformation defines Euler
parameters as functions of Euler angles. We choose here $F({\rho})=\sqrt{\rho}$. This transformation is well
known in the literature (see, for instance, \cite{Ike,Sti,Bar,Cor,Kib84a}). 
The canonical extension associated to the transformation (\ref{parametros}) is  obtained as a
Mathieu transformation, which satisfies  $\sum Q_idq_i= P\,d\rho+\Phi\, d\phi+\Theta\, d\theta+\Psi\,
d\psi$. The relations among the momenta are given by
\begin{eqnarray}\label{momentos}
&& \hspace{-0.3cm}P = \frac{1}{2\sum q_i^2}(q_1Q_1+q_2Q_2+q_3Q_3+q_4Q_4),\nonumber\\[0.8ex]
&&\hspace{-0.3cm}\Theta=\frac{(q_1Q_1+q_2Q_2)(q_3^2 +q_4^2) -(q_3Q_3+q_4 Q_4)(q_1^2 +q_2^2)}
{2\sqrt{(q_1^2 + q_2^2)(q_3^2+q_4^2)}},\nonumber\\
&&\hspace{-0.3cm}\Phi=\frac{1}{2}(-q_2Q_1+q_1Q_2+q_4Q_3-q_3Q_4),\\
&&\hspace{-0.3cm}\Psi =\frac{1}{2}(q_2Q_1-q_1Q_2 +q_4Q_3-q_3Q_4).\nonumber
\end{eqnarray}
Note that the factor $\sqrt{\rho}$ has a long history. Indeed, as  
Bartsch \cite{Bart} put it: `For the one-dimensional Kepler motion, it was already found 
by Euler \cite{Euler}
that  the introduction of a square-root coordinate
$u=\sqrt{x}$ and a fictitious time $\tau$ defined  by $dt=x\,d\tau$ reduces the Kepler 
equation of motion to $d^2u/d\tau^2 +2E\,u=0$ {\it i.e.} the equation of motion of a one-dimensional  harmonic
oscillator'. This reemerges in Heggie and Hut \cite{Heggie} (pp.~145) who seem unaware of this work of Euler.\par

Then, excluding the invariant manifolds ${\cal M}_1\!\!=\{(q,Q)|q_1\!=\!q_2\!=\!0\}$ and ${\cal M}_2=\{(q,Q)|q_3\!=\!q_4\!=\!0\}$ where Levi-Civita transformation already shows the connection of the 2-D  Kepler and isotropic oscillators,  the Hamiltonian (\ref{O4G}) in the new variables may be written as
\begin{eqnarray}\label{newham}
&&\hspace{-0.8cm}{\cal H}_\omega={\cal H}(\rho,\theta,-,-,P,\Theta,\Phi,\Psi)\nonumber\\
&&\hspace{-0.2cm}=  \frac{\rho\, \omega}{2} +2\rho P^2 +
\frac{2}{\rho}\!\!\left(\!\!\Theta^2  +
\frac{\Phi^2+\Psi^2-2\,\Phi\Psi\,\cos\theta}{\sin^2\theta} \right)
\end{eqnarray}
{\it i.e.} variables $\phi$ and $\psi$ are cyclic, with
$\Phi$ and $\Psi$ as the corresponding first integrals. Note that a dash is used instead of the variable  to
stress the fact that this coordinate is ignorable. In other words the differential system  reduces to
\[\frac{d\rho}{d\tau}=\frac{\partial {\cal H}}{\partial P},\quad 
\frac{d\theta}{d\tau}=\frac{\partial {\cal H}}{\partial \Theta},\quad
\frac{dP}{d\tau}=-\frac{\partial {\cal H}}{\partial \rho},\quad
\frac{d\Theta}{d\tau}=-\frac{\partial {\cal H}}{\partial \theta}\]
and two quadratures 
\begin{equation}\label{doscuadraturas}
\phi=\int (\partial {\cal H}/\partial \Phi)\,d\tau \quad {\rm and} \quad 
\psi=\int(\partial {\cal H}/\partial \Psi)\,d\tau.
\end{equation}
In order to reach the main result of this letter, let us remember first
two basic features of our system, presented in the following
Propositions, based on the reordering of the terms defining Eq.
(\ref{newham}) and the suitable chosen regularizing function.
\par\noindent 
{\bf Proposition~1.-} {\sl The 4-D isotropic oscillator may be considered as
a generalized Kepler system. It reduces to a 3-D Kepler system and a quadrature when we restrict to either 
$\Phi=0$ or $\Psi=0$ manifolds.}\medskip\par\noindent 
{\sl Proof}.- Let us consider Poincar\'e technique with a  time regularization
$\tau\rightarrow s$  given  by 
\begin{equation}\label{regular1}
d\tau=(4\rho)^{-1}\,ds.
\end{equation}
Then, the flow is defined by the Hamiltonian ${\tilde{\cal H}}=\frac{1}{4\rho}({\cal H}- h)$ where
$h$ is a fix value of the Hamiltonian ${\cal H}$ for chosen initial conditions, and the flow is
defined now on the manifold ${\tilde{\cal H}}=0$. In a slight different form, it is more
convenient to write
\begin{equation}\label{ham3}
{\tilde{\cal K}}=\frac{1}{2}\left(P^2 + \frac{\Theta^2}{\rho^2}+\frac{\Phi^2
+\Psi^2-2\,\Phi\Psi\,\cos\theta}{\rho^2\,\sin^2\theta}
\right)- \frac{h}{4\rho}
\end{equation}
in the manifold ${\tilde{\cal K}}=-\frac{\omega}{8}$. \par
 Therefore, denoting ${\cal H}_K$ part of function (\ref{ham3}):
\begin{equation}\label{KeplerCoulomb}
{\cal H}_K=\frac{1}{2}\left(P^2 + \frac{\Theta^2}{\rho^2}+\frac{\Phi^2}{\rho^2\,\sin^2\theta}
\right)- \frac{h/4}{\rho},
\end{equation}
then, the Hamiltonian function  (\ref{ham3}) may be also written as
\begin{equation}\label{hamkeplerbis}
{\tilde{\cal K}}= {\cal H}_K + \frac{\Psi^2-2\Phi\Psi\cos \theta}{2\rho^2\,\sin^2\theta}.
\end{equation}
Note that the  function ${\cal H}_K$ is the Hamiltonian of the Kepler system (see \cite{gold}) in spherical coordinates
\begin{equation}\label{esfericascarte}
 x=\rho\,\sin\theta\,\cos\phi, \, \, y=\rho\,\sin\theta\,\sin\phi, \,\, z=\rho\,\cos\theta 
\end{equation}
(where $\theta$ is the colatitude) and their momenta $(P,\Theta,\Phi)$, if we choose
\[\gamma= h/4, \]
where $\gamma$ is the fundamental constant of the Kepler-Coulomb system.
  Observe that we might have  taken the term ${\Psi^2}/({\rho^2\sin^2\theta})$ instead of
${\Phi^2}/({\rho^2\sin^2\theta})$ for the definition of ${\cal H}_K$.\par

The differential system defined by (\ref{hamkeplerbis}) is given by
\begin{eqnarray}\label{ecuaciones}
&&\frac{d\rho}{ds}=\phantom{-}\frac{\partial {\tilde{\cal K}}}{\partial P}=\phantom{-}
\frac{\partial {\cal H}_K}{\partial P},\nonumber\\[0.5ex]
&&\frac{d\theta}{ds}=\phantom{-}\frac{\partial {\tilde{\cal K}}}{\partial \Theta}=\phantom{-}\frac{\partial {\cal
H}_K}{\partial \Theta},\\[0.5ex]
&&\frac{dP}{ds}=-\frac{\partial {\tilde{\cal K}}}{\partial \rho}=-\frac{\partial {\cal
H}_K}{\partial \rho}+\frac{\Psi(\Psi-2\Phi\,\cos \theta)}{\rho^3\,\sin^2\theta},\nonumber\\ [0.5ex]
&&\frac{d\Theta}{ds}=-\frac{\partial {\tilde{\cal K}}}{\partial \theta}=-\frac{\partial {\cal
H}_K}{\partial \theta}- \frac{\Psi(\Psi\cos \theta-\Phi(1+\cos^2 \theta))}{\rho^2\sin^3\theta},\nonumber
\end{eqnarray}
and the two quadratures (\ref{doscuadraturas}). Then, if we restrict to the manifold $\Psi=0$ and we identify the
variable $s$ with the physical time $t$, the flow of the oscillator  given by Eqs. (\ref{ecuaciones}) joint with the 
quadrature $ \phi=\int (\partial {\cal H}_K/\partial \Phi)\,ds$, reduces to a Keplerian flow in
the manifold ${\cal H}_K=-\frac{\omega}{8}$. The last equation is the quadrature of $\psi$ 
\[\psi=\int \frac{\partial \tilde{\cal K}}{\partial \Psi}\,ds=  -\int \frac{\Phi \cos \theta}{\rho^2\sin^2\theta} \,ds , \]
to be computed after the Kepler system is integrated.\hfill  {\bf q.e.d.}\par 
This shows that the 4-D oscillator may be seen as a generalized Kepler system. In fact, in the
literature this Hamiltonian relates to Hartmann and other ring-shaped potentials \cite{Kib87}, 
but we will not enter
that issue here. \par 

Note that, considering the inverse of the Projective Euler transformation (\ref{parametros}), we may see the transformation 
from spherical to Cartesian (\ref{esfericascarte}) as a projection $R^4\rightarrow R^3$. Explicitly, inverting 
(\ref{parametros}) we have $\rho = \sum q_i^2$ and 
\begin{eqnarray}\label{Eulerinversa}
&&\sin\theta =\frac{2\Delta}{\rho}, \quad \cos\theta =\frac{q_3^2 +q_4^2-
q_1^2 -q_2^2 }{\rho},\nonumber\\
&&\sin\phi = \frac{q_1q_3+q_2q_4}{\Delta},
\quad \cos\phi = \frac{q_1q_4-q_2q_3}{\Delta},\nonumber\\
&&\sin\psi = \frac{q_1q_3-q_2q_4}{\Delta},
\quad \cos\psi = \frac{q_1q_4+q_2q_3}{\Delta},\nonumber
\end{eqnarray}
where $\Delta=\sqrt{(q_1^2 +q_2^2)(q_3^2 +q_4^2)}$,  we obtain immediately
\begin{eqnarray*}
&&x=2(q_1q_4-q_2q_3),\\
&&y=2(q_1q_3+q_2q_4),\\
&&z= q_3^2 +q_4^2-q_1^2-q_2^2,
\end{eqnarray*}
in other words, the KS-transformation.
\section{Switching to  Projective Andoyer  variables} 
The Andoyer variables (introduced by Serret \cite{Ser}, and also referred as Serret-Andoyer \cite{Dep93} or Andoyer-Deprit \cite{benettin}) are a well known
symplectomorphism in dynamical astronomy \cite{And22, Dep67, Kin72}  and recently introduced in other fields such 
as attitude and control \cite{Lum}. The reader ought to be aware that in the  literature  authors use
different letters for them (see \cite{And22,Dep67,Boi}). In what follows   the  $(\lambda,\mu,\nu,\Lambda, M,N)$ notation is adopted.

Assuming the vector $(\Phi,\Theta,\Psi)$ different from zero, {\it i.e.} excluding the invariant manifold of rectilinear solutions treated before,  the canonical transformation from  Andoyer  
$(\lambda,\mu,\nu,\Lambda, M,N)$  to Euler $(\phi,\theta,\psi,\Phi,\Theta,\Psi)$, is given by
\begin{eqnarray}\label{ESA6}
&&\cos\epsilon\cos\sigma  - \sin\epsilon\sin\sigma\cos\mu -\cos\theta=0, \nonumber \\
&&\cos\theta\cos\sigma + \sin\theta\sin\sigma\cos\,(\psi-\nu)-\cos\epsilon=0, \nonumber \\
&&\cos\theta\cos\epsilon + \sin\theta \sin\epsilon\cos\,(\phi-\lambda) -\cos\sigma=0,\nonumber\\
&&\Phi= \Lambda ,\\
&& \Psi= N,\nonumber\\
&& \Theta= \sqrt{M^2  - \frac{\Lambda^2 + N^2 - 2N\Lambda\cos\theta}{\sin^2\theta}}.\nonumber
\end{eqnarray}
where $\cos \epsilon= \Lambda/M$ and $\cos \sigma= N/M$. 
In fact, the transformation  requires two charts.\par
When we add to them the variables  
$(\rho, P)$ we obtain what we call the `Projective Andoyer' transformation:
\begin{equation}\label{ESA}
 \Big(\begin{array}{cccc}
\rho,&\!\!\phi,&\!\!\theta,&\!\!\psi\\
P,&\!\!\Phi,&\!\!\Theta,&\!\!\Psi\end{array}\Big) \rightarrow
\Big(\begin{array}{cccc}
\rho,&\!\!\lambda,&\!\!\mu,&\!\!\nu\\
P,&\!\!\Lambda,&\!\!M,&\!\!N\end{array}\Big)
\end{equation}
which, versus Projective Euler transformation, this one is not a canonical extension. Note that completing the 
expressions of the momenta as functions of $(q,Q)$, after some computations we obtain
\begin{equation}\label{momentM}
M = \frac{1}{2}\sqrt{\| q\|^2 \| Q\|^2-(q\cdot Q)^2}.
\end{equation}
which shows that when Projective Andoyer variables are not defined: $M=0$, the motion is rectilinear.\par
\noindent
{\bf Proposition 2.-} {\sl In Projective Andoyer variables, the system defined by the 4-D
isotropic oscillator, properly regularized, is separable in two subsystems, one linear in the angle $\mu$ and a quadrature
for the variable $\rho$.}
\par\noindent
{\sl Proof}.-  Expressing our Hamiltonian (\ref{newham}) in the  `Projective Andoyer' (PA) variables it results
\[{\cal H}_\omega={\cal H}(\rho,-,-,-,P,-, M,-)=\frac{\omega\rho}{2} +2\rho P^2 +\frac{2M^2}{\rho}.\]
We see that the three angle variables $(\lambda,\mu,\nu)$  are 
cyclic. In fact, not only the momenta $(\Lambda, N)$ 
but also  the variables $(\lambda,\nu)$ are integrals. \par
Then, fixing a value of the Hamiltonian $h$ and making a change of independent variable $d\tau=(\rho/4)ds$ the Hamiltonian
takes the form
\[{\cal K}=\frac{\rho}{4}({\cal H}_\omega-h)=\frac{\omega\rho^2}{8}+ \frac{\rho^2P^2}{2}-\frac{h\,\rho}{4} + \frac{1}{2}M^2\]
in the manifold ${\cal K}=0$. From this Hamiltonian we obtain immediately the conclusions. Details are not needed for
our purposes.\hfill  {\bf q.e.d.}\par
Now we present the main result of this letter:\par
\noindent
{\bf Theorem.-} {\sl In Projective Andoyer variables, the system defined by (\ref{O4G}), properly regularized,
includes the Keplerian system for any value of the integral $N$.}
\par\noindent
{\sl Proof}.-
After a change of independent variable according to Poincar\'e technique, ${\cal K} = g(\rho) ({\cal H}_{\omega}-h)$ 
considering now (\ref{regular1}), i.e. $g(\rho) = 1/(4\rho)$, the Hamiltonian of the 4-D isotropic oscillator is
given by 
\begin{equation}\label{ESAT}
\tilde{\cal K} = \frac{1}{2}\Big(P^2 + \frac{M^2}{\rho^2} \Big) - \frac{\gamma}{\rho}
\end{equation}
in the manifold $ \tilde{\cal K} = -  \omega/8$.  
 Now, what remains is to connect this Hamiltonian with the Kepler system in 3-D. The use of the 
polar-nodal 
canonical transformation (see Deprit \cite{Dep81}) extended to four dimensions
\begin{equation}\label{transformpolarnodal}
 (\rho,\lambda,\mu,\nu, P,\Lambda, M,N)\rightarrow (x,y,z,\nu,X,Y,Z,N)
\end{equation}
is one way to do it. Indeed, denoting ${\cal R}({\bf v},\alpha)$  a  rotation matrix of angle $\alpha$ around
the vector ${\bf v}$, the transformation (\ref{transformpolarnodal}) is defined  considering three rotations related to the
three direct orthogonal reference frames
$(e_1,e_2,e_3)$, $(\ell_1,\ell_2,\ell_3)$ and $(b_1,b_2,b_3)$, where 
$\ell_1= {\cal R}(e_3,\lambda)e_1$, $b_3= {\cal R}(\ell_1,I)e_3$ and $b_1= {\cal R}(b_3,\mu)\ell_1$,
by
\begin{eqnarray*}
(x,y,z)^T&\!\! =\!\!&{\cal R}(e_3,\lambda)\,{\cal R}(\ell_1,I)\,{\cal R}(b_3,\mu)\,(\rho,0,0)^T\\
(X,Y,Z)^T&\!\!=\!\!&{\cal R}(e_3,\lambda)\,{\cal R}(\ell_1,I)\,{\cal R}(b_3,\mu)\,(P,\frac{M}{\rho},0)^T,
\end{eqnarray*}
with  $\cos I= {\Lambda}/{M}$, and where $T$ stands for transpose of a vector. Explicitly,
we have
\begin{eqnarray*}
&&x =\rho(\cos \mu\cos \lambda - \sin \mu \sin \lambda\cos I),\\[1ex] 
&&y = \rho(\cos \mu\sin \lambda + \sin \mu \cos \lambda\cos I),\\[1ex]
&&z = \rho \sin \mu\sin I,
\end{eqnarray*}\noindent
and similarly for the momenta. It is easy to verify that (\ref{transformpolarnodal}) is a canonical   
transformation of Mathieu type:
$Xdx+Ydy+Zdz =Pd\rho+\Lambda d\lambda+Md\mu$, but not a canonical extension. Perhaps for this reason is not so well known. 
Then, our Hamiltonian (\ref{ESAT}) expressed the variables $(x,y,z,\nu,X,Y,Z, N)$  
takes the form
\begin{equation}
 {\cal H}= {\cal H}(x,y,z,-,X,Y,Z, -)=\frac{1}{2}\|X\|^2-\frac{\gamma}{\| x\|} 
\end{equation}
which is the Hamiltonian of the 3-D Keplerian system where, with some abuse of notation, we have written $x\equiv (x,y,z)$ and $X\equiv(X,Y,Z)$. Moreover,  $(\nu, N)$ are integrals that take
any value. \hfill  {\bf q.e.d.} \par

Let give details of the inverse process: Let consider a reference frame $({e}_1,{ e}_2,{ e}_3)$. We define the vectors $u$ and $n$ such that  
 $x=\rho u$, where  $\rho=\|x\|$ and  $\|u\|=1$, and ${\bf M}=x\times X = M n$, with $M=\|{\bf M}\|$, $\|{ n}\|=1$. Then, we take the  momenta $P,\Lambda$ as $P=x\cdot X/\|x\|$ and  $\Lambda=xY-yX$. Moreover, we may write 
${e}_3\times {n} = (\sin \lambda) \,{\ell}$, with  $\|\ell\|=1$, and
$\cos \lambda= {e}_1 \cdot \ell$, join with $\cos\mu = \ell\cdot { u}$, and $\sin\mu = ({n}\times\ell)\cdot {u}$, which ends the inversion of the polar nodal transformation. Then from  Eqs.~(\ref{ESA6}) we obtain
the projective Euler  canonical variables. Finally, from  Eqs.~(\ref{parametros}) we obtain $q_i$. It rests to have the
expression for $Q_i$ ($i=1,4$). From the canonical extension  Eqs.~(\ref{momentos}) we know they are linear in them. Thus, we
obtain  explicitly $Q_i$ by inverting the matrix associated with the transformation, but it is not necessary to be given
here. \par 

We see that, comparing with KS and in contrast with it,   no constraint is needed for the integrals $(\nu,N)$.  Then, what gives KS transformation such a special place in the oscillator Kepler connection? The fact that Projective Andoyer transformation is not defined when $M=0$, which according to Eq.~(\ref{momentM}) corresponds to rectilinear trajectories: $q\,||\,Q$.\par

Note that, although the Projective Andoyer variables do not yet define a set of action angle variables, they are an intermediary step in that direction. Indeed, from (\ref{ESAT}) we readely obain a {\sl Delaunay set of variables} for 4-D oscillators, a symplectomorphism well suited for perturbation theories. We just need to make use of the Delaunay transformation $(\rho,\mu,P,M)\rightarrow (\delta,g,D,G)$
(see \cite{gold},\cite{Dep81}). It reduces Eq. (\ref{ESAT}), {\it i.e.} (\ref{O4G}), to only one action ${\cal H}=-\gamma^2/(2D^2)$ (for details \cite{Fer}),
which reflects the fact that our systems are maximally superintegrable \cite{Fasso}.\par
\medskip
\par\noindent
\hfill
\hbox{$
\begin{array}[t]{ccc}
\Big(\!\!\begin{array}{cccc}
q_1,&\!\!q_2,&\!\!q_3,&\!\!q_4\\
Q_1,&\!\!Q_2,&\!\!Q_3,&\!\!Q_4
\end{array}\!\!\Big) & 
 \stackrel{\rm Projective\,\,Euler }{\longrightarrow} & 
 \Big(\begin{array}{cccc}
\rho,&\!\!\phi,&\!\!\theta,&\!\!\psi\\
P,&\!\!\Phi,&\!\!\Theta,&\!\!\Psi\end{array}\Big)\\ 
\Biggm\downarrow{} &  & \Biggm\downarrow{{\rm P\,\!A}} \\ 
\Big(\begin{array}{cccc}
x,&\!\!y,&\!\!z,&\!\!\nu\\
X,&\!\!Y,&\!\!Z,&\!\!N
\end{array}\Big)  &
\stackrel{\rm Polar\,\, Nodal}{\longleftarrow}   & 
\Big(\begin{array}{cccc}
\rho,&\!\!\lambda,&\!\!\mu,&\!\!\nu\\
P,&\!\!\Lambda ,&\!\!M,&\!\! N\end{array}\Big)
\end{array}$}
\hfill
\medskip
\par\noindent


This diagram, based on the group structure of canonical transformations, suggests to carry out explicit
compositions of some of them, just to obtain more insight, as well as possible relations with other known
transformations, or new ones that might be defined likewise. 
Further relations among the previous transformations, the option $F(\rho)=\rho$ included, as well as the connection with
other  integrable  and perturbed systems is in progress \cite{Fer}. 
\par

\medskip\par
Thanks to Dr. Egea for his help in an early version of this letter and to Dr. 
Lara for his  comments.
Partial support came  from projects MTM2006-06961 and MTM2009-10767 of
the Ministry of Technology and Science of Spain, and a grant from Fundaci\'on S\'eneca, of the Autonomous
Region of Murcia. 

\end{document}